\documentclass{article}

\usepackage{arxiv}

\usepackage[utf8]{inputenc} 
\usepackage[T1]{fontenc}    
\usepackage{hyperref}       
\usepackage{url}            
\usepackage{booktabs}       
\usepackage{amsfonts}       
\usepackage{nicefrac}       
\usepackage{microtype}      
\usepackage{lipsum}

\usepackage{balance}
\usepackage{natbib}
\setcitestyle{authoryear,open={(},close={)}}
\usepackage{placeins}
\usepackage{graphicx}
    
\begin{document}

\title{U.S. Public Opinion on the Governance of Artificial Intelligence}

\author{
 Baobao Zhang and Allan Dafoe\thanks{The research was funded by the Ethics and Governance of Artificial Intelligence Fund and Good Ventures. For useful feedback we would like to thank: Miles Brundage, Jack Clark, Kanta Dihal, Jeffrey Ding, Carrick Flynn, Ben Garfinkel, Rose Hadshar, Tim Hwang, Katelynn Kyker, Jade Leung, Luke Muehlhauser, Cullen O’Keefe, Michael Page, William Rathje, Carl Shulman, Brian Tse, Remco Zwetsloot, and the YouGov Team (Marissa Shih and Sam Luks). In particular, we are grateful for Markus Anderljung’s insightful suggestions and detailed editing. We acknowledge Steven Van Tassell for his copy-editing and Will Marks and Catherine Peng for their research assistance.} \\
  Centre for the Governance of AI\\
  Future of Humanity Institute, University of Oxford\\
Oxford, OX1 1PT United Kingdom \\
  \texttt{baobaozhangresearch@gmail.com} \\
  \texttt{allan.dafoe@politics.ox.ac.uk} \\
}

\maketitle

\begin{abstract}
Artificial intelligence (AI) has widespread societal implications, yet social scientists are only beginning to study public attitudes toward the technology. Existing studies find that the public's trust in institutions can play a major role in shaping the regulation of emerging technologies. Using a large-scale survey ($N$=2000), we examined Americans' perceptions of 13 AI governance challenges as well as their trust in governmental, corporate, and multistakeholder institutions to responsibly develop and manage AI. While Americans perceive all of the AI governance issues to be important for tech companies and governments to manage, they have only low to moderate trust in these institutions to manage AI applications. 
\end{abstract}

\section{Introduction}

\noindent Advances in artificial intelligence (AI) could impact nearly all aspects of society, including the labor market, transportation, healthcare, education, and national security \citep{oecdai}. AI's effects may be profoundly positive, but the technology entails risks and disruptions that warrant attention. While technologists and policymakers have begun to discuss the societal implications of machine learning and AI more frequently, public opinion has not shaped much of these conversations. Given AI's broad impact, civil society groups argue that the public, particularly those underrepresented in the tech industry, should have a role in shaping the technology \citep{west2019}.

In the U.S., public opinion has shaped policy outcomes \citep{caughey2018policy}, including those concerning immigration, free trade, international conflicts, and climate change mitigation. As in these other policy domains, we expect the public to become more influential over time in impacting AI policy. It is thus vital to have a better understanding of how the public thinks about AI and the governance of AI. Such understanding is essential to crafting informed policy and identifying opportunities to educate the public about AI's character, benefits, and risks.

\begin{figure}
\centering
\includegraphics{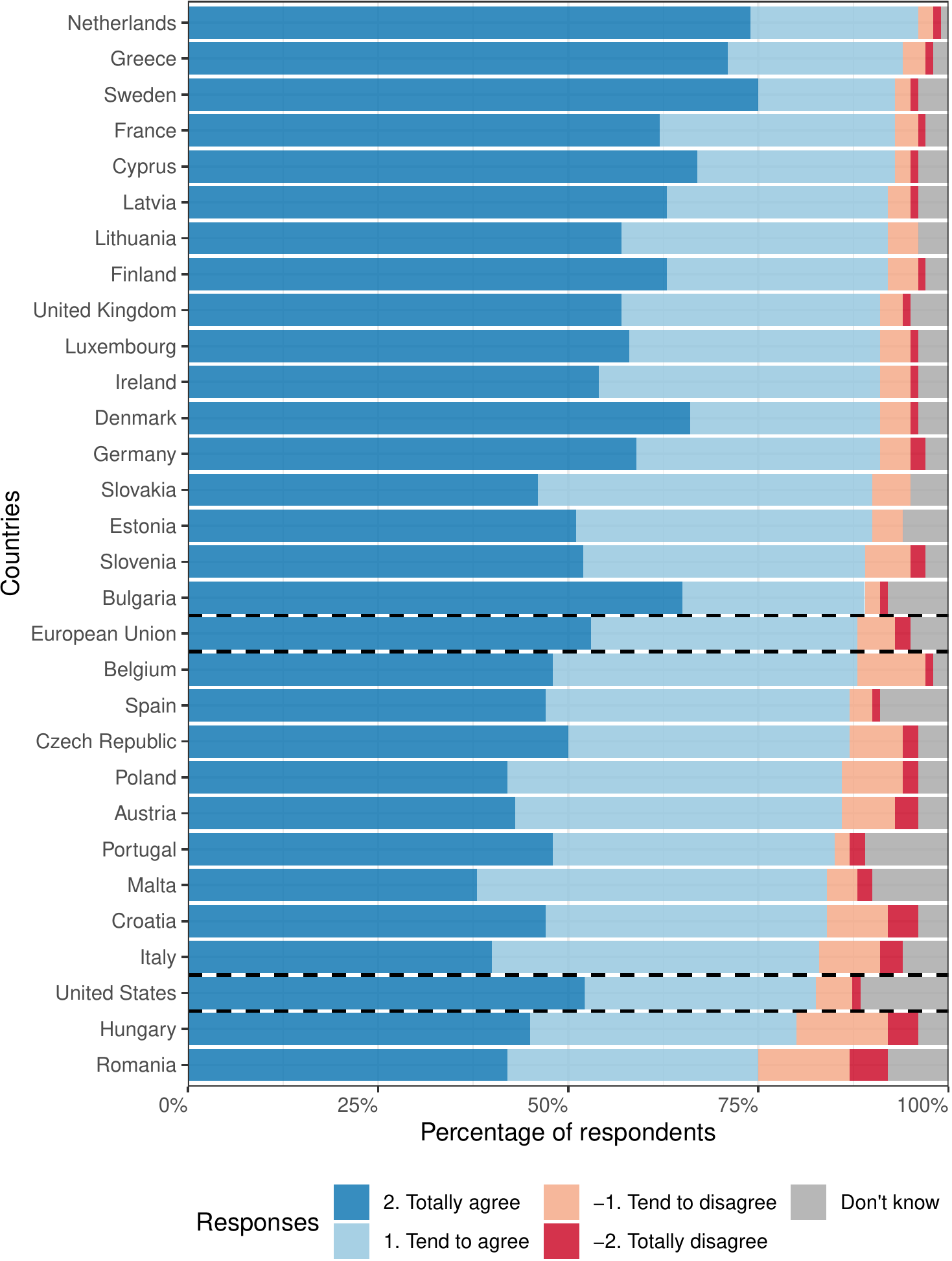}
\caption{Agreement with statement that robots and AI require careful management (EU data from 2017 Special Eurobarometer \#460)}
\label{fig:manage}
\end{figure}

Using an original, large-scale survey ($N$=2000), we studied how the American public perceives AI governance. The overwhelming majority of Americans (82\%) believe that AI and/or robots should be carefully managed. This statistic is comparable to survey results from respondents residing in European Union countries, as seen in Figure \ref{fig:manage}. Furthermore, Americans consider all of the 13 AI governance challenges presented in the survey to be important for tech companies and governments to manage carefully. 

At the same time, Americans have only low to moderate levels of trust in governmental, corporate, and multistakeholder institutions to develop and manage AI in the public's interest. Public trust varies greatly between organizations. Broadly, the public puts the most trust in university researchers (50\% reporting ``a fair amount of confidence'' or ``a great deal of confidence'') and the U.S. military (49\%); followed by scientific organizations, the Partnership on AI, tech companies (excluding Facebook), and intelligence organizations; followed by U.S. federal or state governments, and the UN; followed by Facebook. Contrary to existing research on attitudes toward other emerging technologies, our study finds that individual-level trust in various actors to responsibly develop and manage AI does not predict one's general support for developing AI. 

\section{Background}

\subsection{Public Opinion and AI Governance}

Studying public opinion allows us to anticipate how electoral politics could shape AI governance as policymakers seek to regulate applications of the technology. Over the past two years, corporations, governments, civil society groups, and multistakeholder organizations have published dozens of high-level AI ethics principles. These documents borrow heavily from core principles in bioethics or medical ethics \citep{floridi2019unified}. But unlike medical ethics, which is guided by the common aim of promoting the health of the patient, AI development does not share one common goal but many goals \citep{mittelstadt2019ai}. Furthermore, many of these AI ethics principles are at tension with each other \citep{whittlestone2019role}. For instance, how does one improve the accuracy of algorithmic predictions while ensuring fair and equal treatment of those impacted by the algorithm? As corporations and governments attempt to translate these principles into practice, tensions between principles could lead to political contestation. 

In the U.S., voters are divided on how to regulate facial recognition technology and algorithms used to display social media content, with some of these divisions reflecting existing disagreements between partisans \citep{zhang2019regulation}. For instance, Democrats, compared with Republicans, are much more opposed to law enforcement using facial recognition technology \citep{smith2019facial}. While several states and cities that are heavily Democrat have adopted or proposed moratoriums on law enforcement using facial recognition technology, progress to enact federal regulation has been slow. Industry self-regulation is not immune to partisan politics. For example, Google's AI ethics board dissolved after the company's employees and external civil society groups protested the inclusion of Heritage Foundation president Kay Coles James and drone company CEO Dyan Gibbens on the board. Recognizing the public's divergent policy preferences on AI governance issues is necessary to have a productive public policy deliberation. 

\subsection{How Trust in Institutions Affects Regulation of Emerging Technology}

The public's trust in various institutions to develop and manage AI could affect the regulation of the technology. General social trust is negatively correlated with perceived risk from technological hazards \citep{siegrist2005perception}. One observational study finds that that general social distrust is positively correlated with support for government regulation, even when the public perceives the government to be corrupt \citep{aghion2010regulation}. A follow-up study using more extensive survey data suggests that the public prefers governmental regulation if they trust the government more than they do major companies \citep{pitlik2015does}. This latter finding is supported by how the public has reacted to genetically modified (GM) foods and nanotechnology. These two emerging technologies are similar to AI in that the public has to rely on those with scientific expertise when evaluating potential risks. 

Distrust in institutions producing and regulating GM foods is a compelling explanation for the widespread opposition to GM foods in developed countries. Those with a high-level trust of scientists and regulators are more accepting of GM foods; in contrast, distrust of the agricultural/food industry, contrasted with trust in environmental watchdogs, predicts opposition to GM foods \citep{siegrist1999causal,siegrist2000influence,lang2005does,marques2015attitudes,olofsson2006attitudes}. Although scientists are among the most trusted group in the U.S., Americans have cynical views toward scientists when considering GM foods. Only 19\% thinks that scientists understand the health consequences of GM foods very well, even though scientists have formed a consensus that GM foods are safe to eat. Furthermore, the American public believes that scientists are more motivated by concerns for their industry than concerns for the public \citep{funk2016new}.

Nanotechnology, though less salient than GM foods in the media, is the subject of extensive public opinion research. A meta-analysis of 11 surveys conducted in developed countries finds that the public perceives that the use of nanotechnology has greater benefits than risks; nevertheless, a large subset of the public is uncertain of the consequences of nanotechnology \citep{satterfield2009anticipating}. As in perceptions of GM foods, trust in institutions seems to play a significant role in shaping attitudes toward nanotechnology. Americans who have lower confidence in business leaders within the nanotechnology industry also perceive the technology to be riskier \citep{macoubrie2004public}. A higher level of distrust in government agencies to protect the public from nanotechnology hazards is associated with a higher perceived risk of the technology \citep{macoubrie2006nanotechnology,siegrist2007laypeople}. Consumers who are less trustful of the food industry indicate they are more reluctant to buy foods produced or packaged using nanotechnology \citep{siegrist2007public}. 

\subsection{Existing Survey Research on AI Governance and Trust in Tech Companies}

Our survey builds on existing public opinion research on attitudes toward AI and trust in tech companies. Past survey research related to AI tends to focus on specific governance challenges, such as lethal autonomous weapons \citep{horowitz2016public}, algorithmic fairness \citep{saxena2019fairness}, or facial recognition technology \citep{smith2019facial,facevalue2019}. A few large-scale surveys have taken a more comprehensive approach by asking about a range of AI governance challenges \citep{eurobarometer460,smith2017,cave2019scary}. We improved upon these surveys by asking respondents to consider a variety of AI governance challenges using the same question-wording. This consistency in question-wording allowed us to compare respondents' perceptions toward major issues in AI governance. 

Another innovation of our research project is that we connect AI governance with trust in tech companies. Trust in tech companies has become a growing topic in survey research. While Americans perceive tech companies to have a more positive impact on society than other institutions, including the news media and the government, their feelings toward tech companies have declined dramatically since 2015 \citep{doherty2019americans}. Nevertheless, there exists heterogeneity in how the public views individual tech companies; one consistent finding in non-academic surveys is that the public strongly distrusts Facebook, particularly in its handling of personal data \citep{newton2017verge,molla2018facebook,kahn2018americans}. In our survey, we examined whether this general distrust of tech companies extend to their management of AI development and applications. 

\section{Survey Data and Analysis}

We conducted an original online survey ($N$=2000) through YouGov between June 6 and 14, 2018. YouGov drew a random sample from the U.S. adult population (i.e., the target sample) and selected respondents from its online respondent panel that matched the target sample on key demographic variables. The details of YouGov's sample matching methodology can be found in the Appendix. 

We pre-registered nearly all of the analysis on Open Science Framework (pre-analysis plan URL: \url{https://osf.io/7gqvm/}). Pre-registration increases research transparency by requiring researchers to specify their analysis before analyzing the data \citep{nosek2018preregistration}. Doing so prevents researchers from misusing data analysis to come up with statistically significant results when they do not exist, otherwise known as $p$-hacking. Survey weights provided by YouGov were used in our primary analysis. We followed the Standard Operating Procedures for Don Green's Lab at Columbia University when handling missing data or ``don't know'' responses \citep{lin2016standard}. Heteroscedasticity-consistent standard errors were used to generate the margins of error at the 95\% confidence level. We report cluster-robust standard errors whenever there is clustering by respondent. In figures, each error bar shows the 95\% confidence intervals. 

\section{Perceptions of AI Governance Challenges}

\subsection{Methodology}

We sought to understand how Americans prioritize policy issues associated with AI. Respondents were asked to consider five AI governance challenges, randomly selected from a set of 13 (see the Appendix for the full text); the order these five were given to each respondent was also randomized. After considering each governance challenge, respondents were asked how likely they think the challenge will affect large numbers of people both 1) in the U.S. and 2) around the world within 10 years using a seven-point scale that displayed both numerical likelihoods and qualitative descriptions (e.g., ``Very unlikely: less than 5\% chance (2.5\%)''). Respondents were also asked to evaluate how important it is for tech companies and governments to carefully manage each challenge presented to them using a four-point scale.

\subsection{General Results}

\begin{figure}
\centering
\includegraphics[height=0.45\textheight]{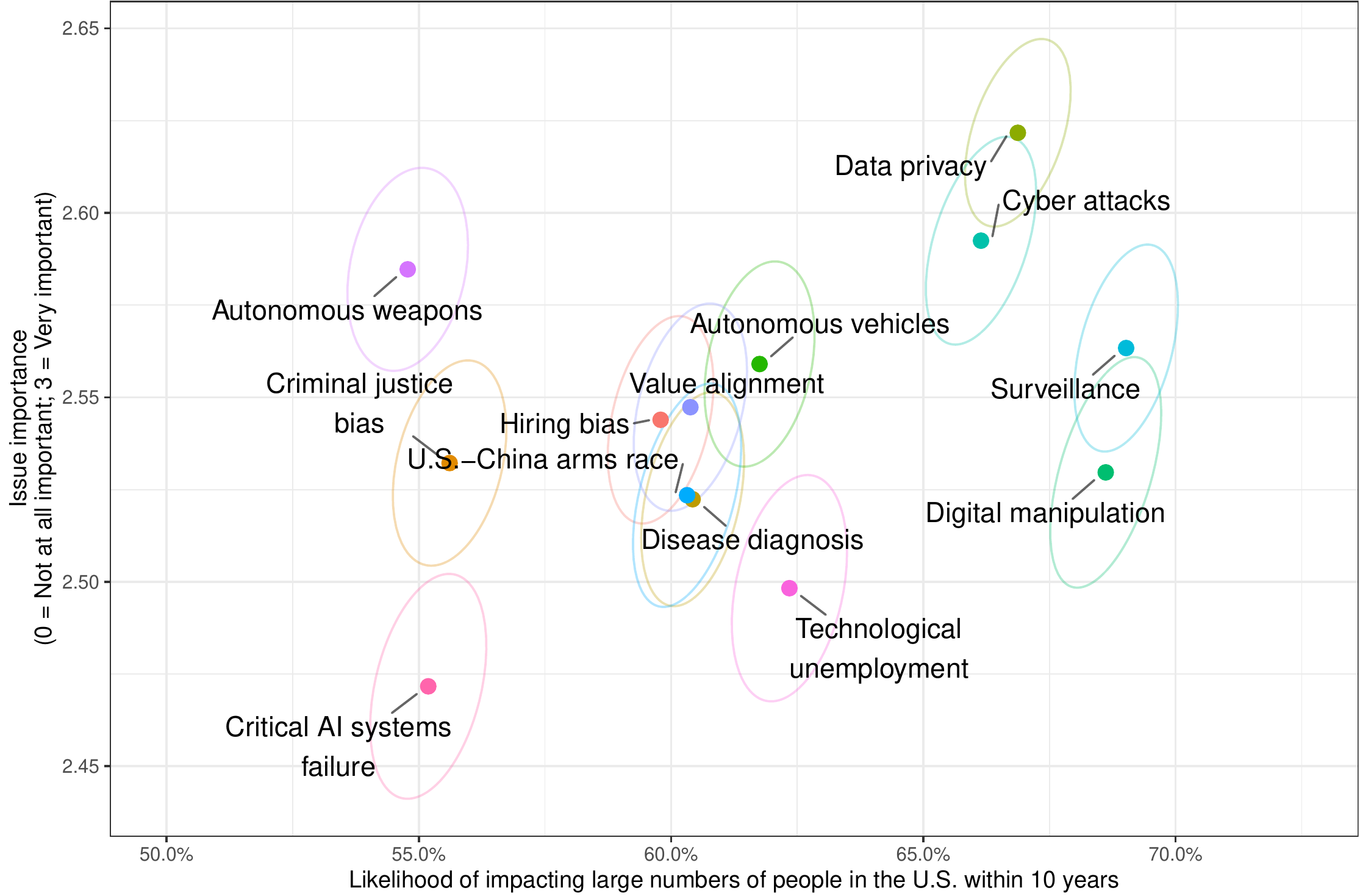} 
\includegraphics[height=0.45\textheight]{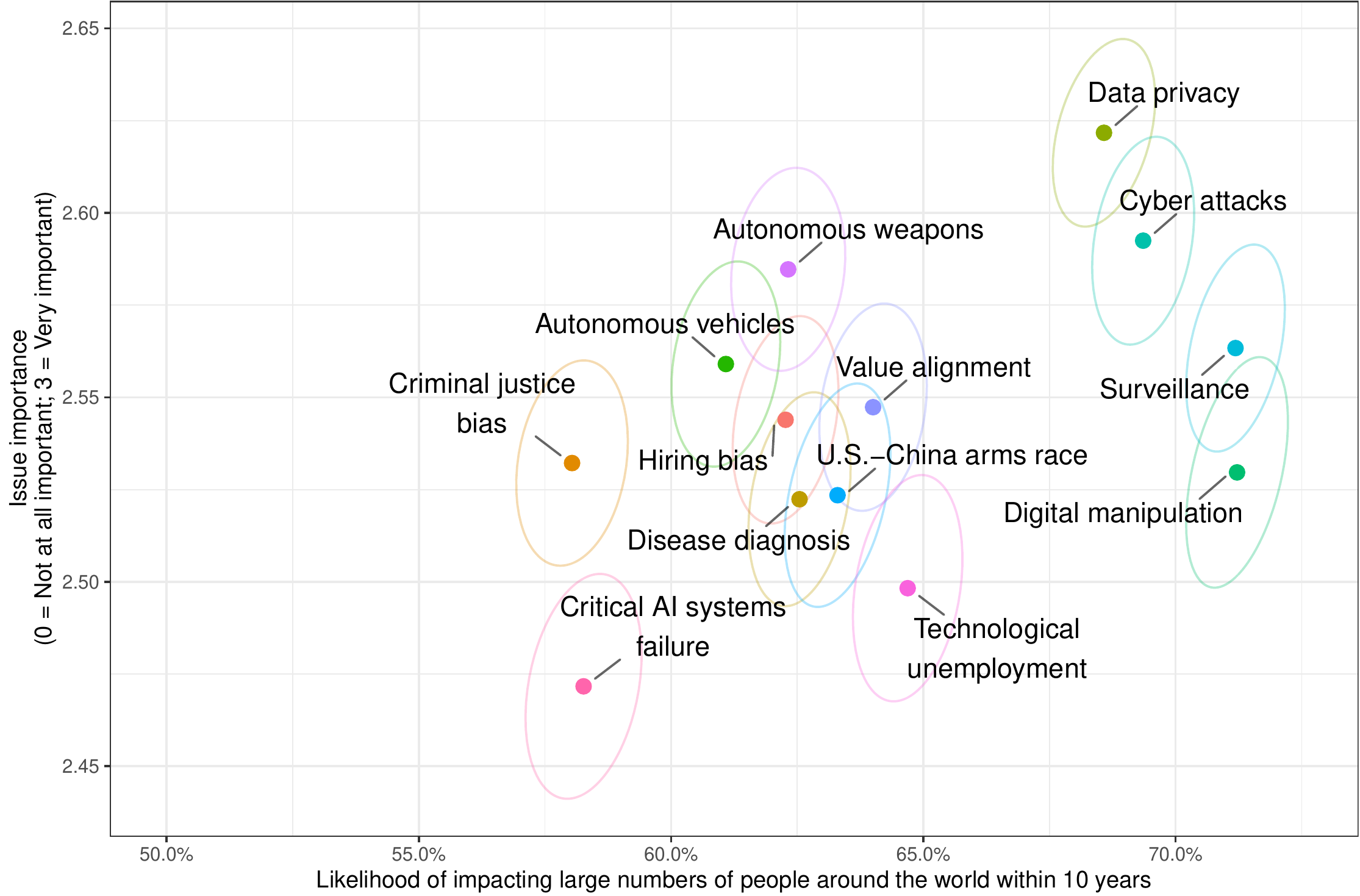}
\caption{Perceptions of AI governance challenges in the U.S. and around the world}
\label{fig:aigovchallenges}
\end{figure}

We use scatterplots to visualize our survey results. In the top graph in Figure \ref{fig:aigovchallenges}, the \emph{x}-axis is the perceived likelihood of the problem happening to large numbers of people in the U.S. In the bottom graph, the \emph{x}-axis is the perceived likelihood of the problem happening to large numbers of people around the world. The \emph{y}-axes on both graphs in Figure \ref{fig:aigovchallenges} represent respondents' perceived issue importance, from 0 (not at all important) to 3 (very important). Each dot represents the mean perceived likelihood and issue importance. The correspondent ellipse represents the 95\% confidence region of the bivariate means assuming the two variables are distributed multivariate normal.

Americans consider all the AI governance challenges we present to be important: the mean perceived issues importance of each governance challenge is between ``somewhat important'' (2) and ``very important'' (3), though there is meaningful and discernible variation across items.

The AI governance challenges Americans think are most likely to impact large numbers of people, and are important for tech companies and governments to tackle, are found in the upper-right quadrant of the two plots. These issues include data privacy as well as AI-enhanced cyber attacks, surveillance, and digital manipulation. We note that the media have widely covered these issues during the time of the survey.

There are a second set of governance challenges that are perceived on average, as about 7\% less likely, and marginally less important. These include autonomous vehicles, value alignment, bias in using AI for hiring, the U.S.-China arms race, disease diagnosis, and technological unemployment. Finally, the third set of challenges are perceived on average another 5\% less likely, and about equally important, including criminal justice bias and critical AI systems failures.

We also note that Americans predict that all of the governance challenges mentioned in the survey, besides protecting data privacy and ensuring the safety of autonomous vehicles, are more likely to impact people around the world than to affect people in the U.S. While most of these statistically significant differences are substantively small, one difference stands out: Americans think that autonomous weapons are 7.6 percentage points more likely to impact people around the world than Americans (two-sided \(p\)-value \textless{} 0.001).

We want to reflect on one result. ``Value alignment'' consists of an abstract description of the alignment problem and a reference to what sounds like individual level harms: ``while performing jobs they could unintentionally make decisions that go against the values of its human users, such as physically harming people.'' ``Critical AI systems failures,'' by contrast, references military or critical infrastructure uses, and unintentional accidents leading to ``10 percent or more of all humans to die.'' The latter was weighted as less important than the former: we interpret this as a probability-weighted assessment of importance, since presumably the latter, were it to happen, is much more important. We thus think the issue importance question should be interpreted in a way that down-weights low probability risks. 

\subsection{Subgroup Analysis}

\begin{figure}
\centering
\includegraphics[width=0.90\textwidth]{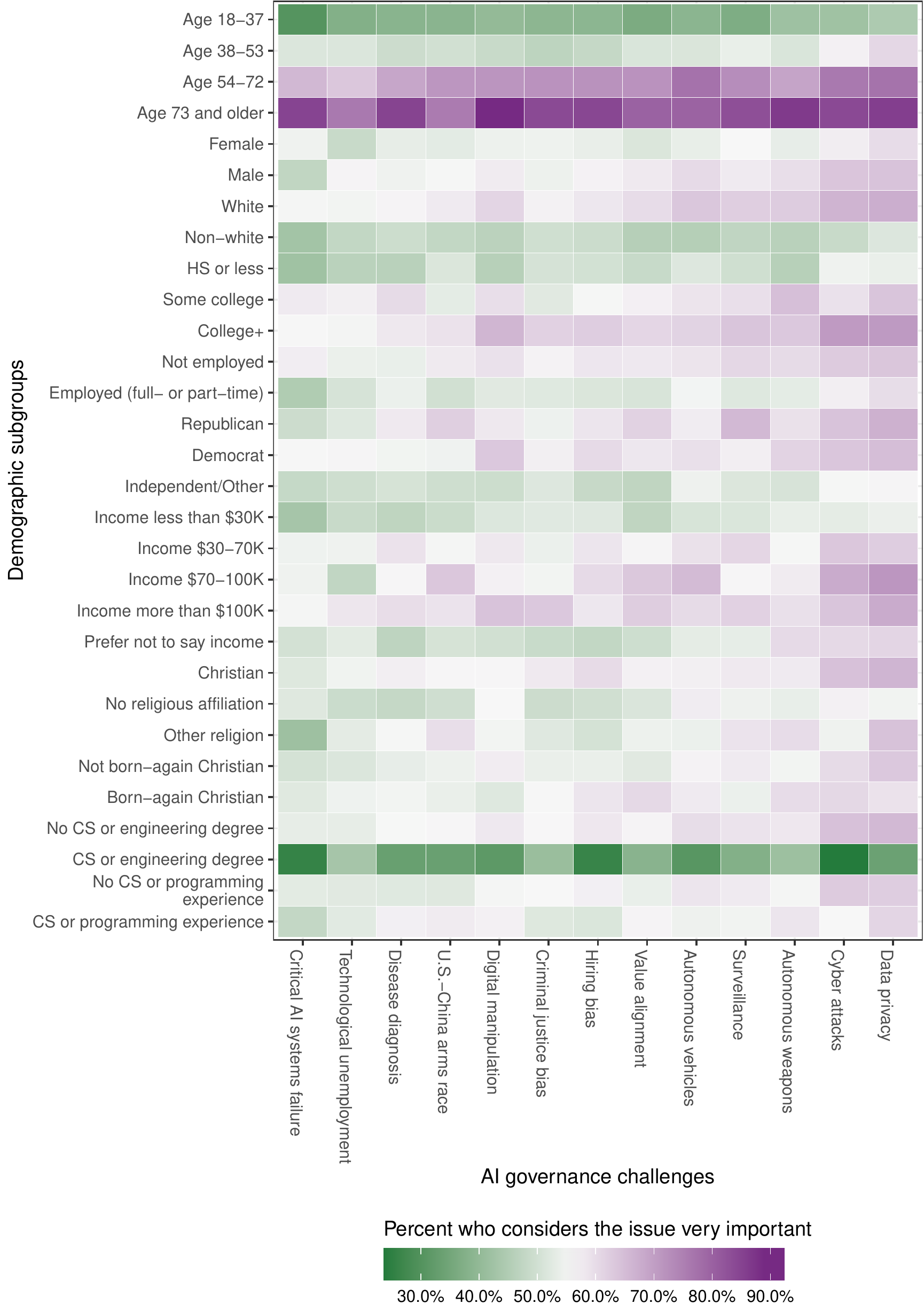}
\caption{AI governance challenges: issue importance by demographic subgroups}
\label{fig:subgroup}
\end{figure}

We performed further analysis by calculating the percentage of respondents in each subgroup who consider each governance challenge to be ``very important'' for governments and tech companies to manage. In general, differences in responses are more salient across demographic subgroups than across governance challenges. In a linear multiple regression predicting perceived issue importance using demographic subgroups, governance challenges, and the interaction between the two, we find that the stronger predictors are demographic subgroup variables, including age group and having CS or programming experience.

Two highly visible patterns emerge from our data visualization. First, a higher percentage of older respondents, compared with younger respondents, consider nearly all AI governance challenges to be ``very important.'' In another part of the survey (see Figure 7 in the Appendix), we find that older Americans, compared with younger Americans, are less supportive of developing AI. Our results here might explain this age gap: older Americans see each AI governance challenge as substantially more important than do younger Americans. Whereas 85\% of Americans older than 73 consider each of these issues to be very important, only 40\% of Americans younger than 38 do.

Second, those with CS or engineering degrees, compared with those who do not, rate all AI governance challenges as less important. This result could explain another finding in our survey that shows those with CS or engineering degrees tend to express greater support for developing AI. In Table 4 in the Appendix, we report the results of a saturated linear model using demographic variables, governance challenges, and the interaction between these two types of variables to predict perceived issue importance. We find that those who are 54-72 or 73 and older, relative to those who are below 38, view the governance challenges as more important (two-sided \(p\)-value \textless{} 0.001 for both comparisons). Furthermore, we find that those who have CS or engineering degrees, relative to those who do not, view the governance challenges as less important (two-sided \(p\)-value \textless{} 0.001).

\section{Trust in Actors to Develop and Manage AI}

\subsection{Methodology}

Respondents were asked how much confidence they have in various actors to develop AI. They were randomly assigned five actors out of 15 to evaluate. We provided a short description of actors that are not well-known to the public (e.g., NATO, CERN, and OpenAI). Also, respondents were asked how much confidence, if any, they have in various actors to manage the development and use of AI in the best interests of the public. They were randomly assigned five out of 15 actors to evaluate. Again, we provided a short description of actors that are not well-known to the public (e.g., AAAI and Partnership on AI). Confidence was measured using the same four-point scale described above. The two sets of 15 actors differed slightly because, for some actors, it seemed inappropriate to ask one or the other question. See the Appendix for the exact wording of the questions and descriptions of the actors.

\subsection{Results}

\begin{figure}
\centering
\includegraphics[width=0.90\textwidth]{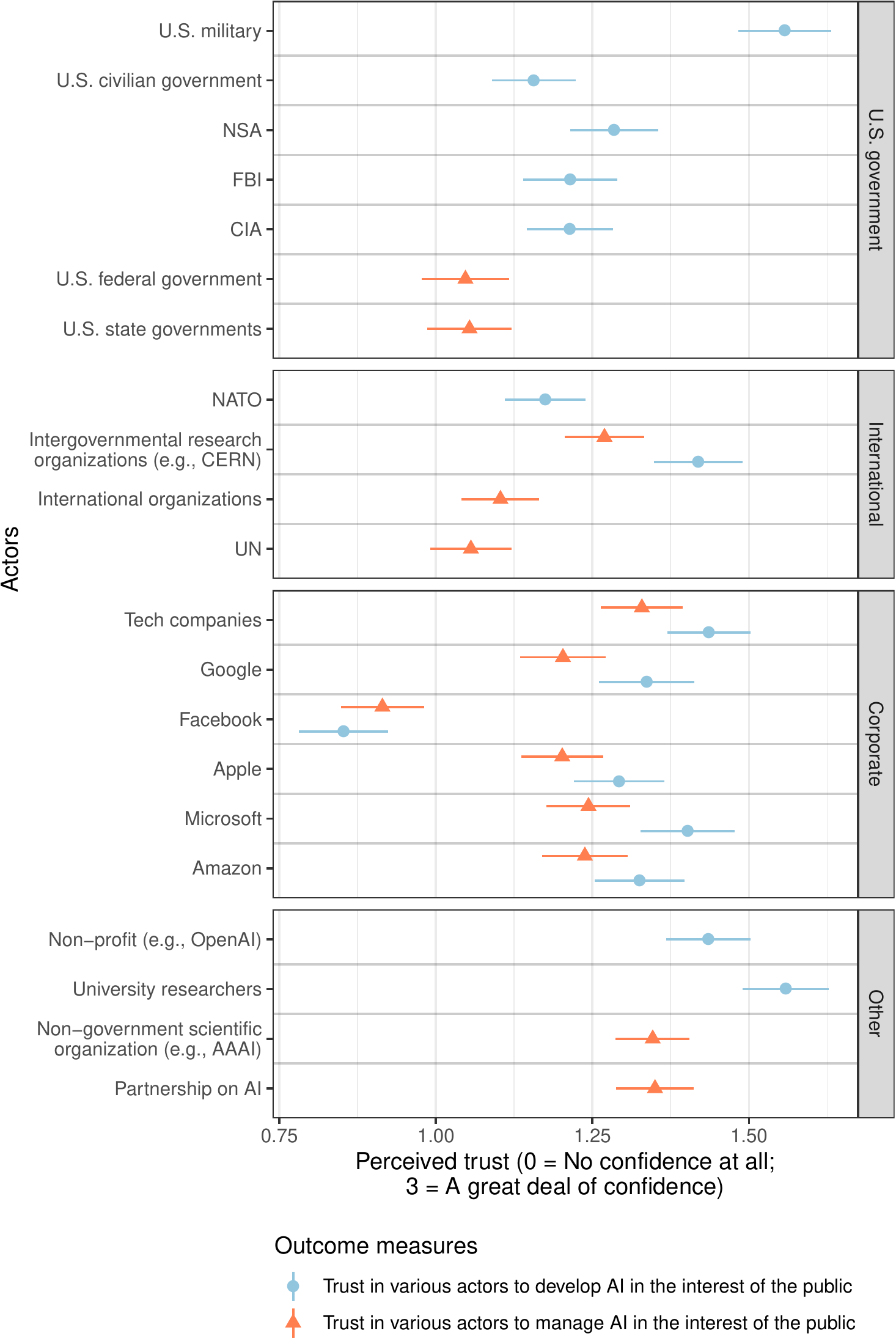}
\caption{Trust in various actors to develop and manage AI in the interest of the public}
\label{fig:trust}
\end{figure}

Americans do not express great confidence in most actors to develop or to manage AI, as seen in Figures 5 and 6 in the Appendix. A majority of Americans do not have a ``great deal'' or even a ``fair amount'' of confidence in any institution, except university researchers, to develop AI. Furthermore, Americans place greater trust in tech companies and non-governmental organizations (e.g., OpenAI) than in governments to manage the development and use of the technology.

University researchers and the U.S. military are the most trusted groups to develop AI: about half of Americans express a ``great deal'' or even a ``fair amount'' of confidence in them. Americans express slightly less confidence in tech companies, non-profit organizations (e.g., OpenAI)\footnote{At the time the survey was conducted (June 2018), OpenAI was a 501(c)(3) nonprofit organization. In March 2019, OpenAI announced it is restructuring into ``capped-profit'' company to attract investments.}, and American intelligence organizations. Nevertheless, opinions toward individual actors within each of these groups vary. For example, while 44\% of Americans indicated they feel a ``great deal'' or even a ``fair amount'' of confidence in tech companies, they rate Facebook as the least trustworthy of all the actors. More than four in 10 indicate that they have no confidence in the company.

Our survey was conducted between June 6 and 14, 2018, shortly after the Facebook/Cambridge Analytica scandal became highly salient in the media. On April 10-11, 2018, Facebook CEO Mark Zuckerberg testified before the U.S. Congress regarding the Cambridge Analytica data leak. On May 2, 2018, Cambridge Analytica announced its shutdown. Nevertheless, Americans' distrust of the company existed before the Facebook/Cambridge Analytica scandal. In a pilot survey that we conducted on Mechanical Turk during July 13--14, 2017, respondents indicated a substantially lower level of confidence in Facebook, compared with other actors, to develop and regulate AI.

The results on the public's trust of various actors to manage the develop and use of AI provided are similar to the results discussed above. Again, a majority of Americans do not have a ``great deal'' or even a ``fair amount'' of confidence in any institution to manage AI. In general, the public expresses greater confidence in non-governmental organizations than in governmental ones. Indeed, 41\% of Americans express a ``great deal'' or even a ``fair amount'' of confidence in ``tech companies,'' compared with 26\% who feel that way about the U.S. federal government. But when presented with individual big tech companies, Americans indicate less trust in each than in the broader category of ``tech companies.'' Once again, Facebook stands out as an outlier: respondents give it a much lower rating than any other actor. Besides ``tech companies,'' the public places relatively high trust in intergovernmental research organizations (e.g., CERN)\footnote{We asked about trust in intergovernmental research organizations, such as CERN, because some policy researchers have proposed creating a politically neutral AI development hub similar to CERN to avoid risks associated with competition in AI development between rival states \citep{politicallyneutral2019}.}, the Partnership on AI, and non-governmental scientific organizations (e.g., AAAI). Nevertheless, because the public is less familiar with these organizations, about one in five respondents give a ``don't know'' response.

Similar to our findings, recent survey research suggests that while Americans feel that AI should be regulated, they are unsure \emph{who} the regulators should be. When asked who ``should decide how AI systems are designed and deployed,'' half of Americans indicated they do not know or refused to answer \citep{west2018divided}. Our survey results also seem to reflect Americans' general attitudes toward public institutions. According to a 2016 Pew Research Center survey, an overwhelming majority of Americans have ``a great deal'' or ``a fair amount'' of confidence in the U.S. military and scientists to act in the best interest of the public. In contrast, public confidence in elected officials is much lower: 73\% indicated that they have ``not too much confidence'' or ``no confidence'' \citep{funk2017}. Less than one-third of Americans thought that tech companies do what's right ``most of the time'' or ``just about always''; moreover, more than half indicate that tech companies have too much power and influence in the U.S. economy \citep{smith2018}. Nevertheless, Americans' attitudes toward tech companies are not monolithic but varies by company. For instance, our research findings reflect the results from several non-academic surveys that find the public distrusts Facebooks significantly more than other major tech companies \citep{newton2017verge,molla2018facebook,kahn2018americans}.

\subsection{Predicting Support for Developing AI Using Institutional Trust}

\begin{table}[!htbp] \centering 
  \caption{Regression results: predicting support for developing AI using respondents' trust in different types of actors to develop AI} 
  \label{tab:trustdevsupport} 
\begin{tabular}{@{\extracolsep{5pt}} lll} 
\\[-1.8ex]\hline 
\hline \\[-1.8ex] 
Variable & Coefficient (SE) & $p$-value \\ 
\hline \\[-1.8ex] 
Corporate actors & -0.02 (0.03) & 0.596 \\ 
International actors & -0.01 (0.03) & 0.830 \\ 
U.S. government actors & -0.01 (0.03) & 0.799 \\
Intercept & 0.26 (0.03) & \textless 0.001 \\ 
\hline \\[-1.8ex] 
\end{tabular} 

\raggedright
$N$ = 10000 observations from 2000 respondents (standard errors are clustered by respondent); $F$-statistic: 3.976 on 79 and 1999 DF, $p$-value: $<0.001$. We controlled for the demographic variables found in Table 3.
\end{table} 

\begin{table}[!htbp] \centering 
  \caption{Regression result: predicting support for developing AI using respondents' trust in different types of actors to manage AI} 
  \label{tab:trustmanageupport} 
\begin{tabular}{@{\extracolsep{5pt}} lll} 
\\[-1.8ex]\hline 
\hline \\[-1.8ex] 
Variable & Coefficient (SE) & $p$-value \\ 
\hline \\[-1.8ex] 
Corporate actors & -0.03 (0.02) & 0.271 \\ 
International actors & 0.02 (0.03) & 0.517 \\ 
U.S. government actors & 0.02 (0.03) & 0.449 \\ 
Intercept & 0.26 (0.03) & \textless 0.001 \\
\hline \\[-1.8ex] 
\end{tabular} 

\raggedright
$N$ = 10000 observations from 2000 respondents (standard errors are clustered by respondent); $F$-statistic: 3.998 on 79 and 1999 DF, $p$-value: $<0.001$. We controlled for the demographic variables found in Table 3.
\end{table} 

The American public's support for developing AI is not predicted by their trust in various actors to develop and manage AI. Our null results stand in contrast to studies of GM foods and nanotechnology, where trust in institutions that develop and regulate the technology is associated with greater support for the technology. 

For our analysis, we use multiple linear regression to whether individual-level trust in various types of institutions can predict support for developing AI. Note that this analysis is not included in our pre-registration and is therefore exploratory in nature. There are four categories of actors (the same ones shown in Figure \ref{fig:trust}): U.S. government, international bodies, corporate, and others (e.g., universities and non-profits). In our regressions, the group ``others'' is the reference group. Support for developing AI is measured using a five-point Likert scale, with -2 meaning ``strongly oppose'' and 2 meaning ``strongly support.'' Our regressions controlled for all the demographic variables shown in Table 3 in the Appendix. Trust in none of the actor types predicts support for developing AI, as seen in Tables \ref{tab:trustdevsupport} and \ref{tab:trustmanageupport}.

\section{Conclusion}

To understand what the American public thinks about the regulation of the technology, we conducted a large-scale survey. The survey reveals that while Americans consider all AI governance challenges to have high issue importance, they do not necessarily trust the actors who have the power to develop and manage the technology to act in the public's interest. Nevertheless, as our exploratory analysis from the previous section shows, institutional distrust does not necessarily predict opposition to AI development. 

One direction for future research is to examine other factors that shape the public's preferences toward AI governance. Technological knowledge, moral and psychological attributes, and perceptions of risks versus benefits are associated with support for GM foods and nanotechnology \citep{scott2018overview,satterfield2009anticipating}. Another direction for future research is to improve the measurement of institutional trust. Our existing survey questions focused on the public interest component of trust; future studies could investigate other components of trust, such as competence, transparency, and honesty \citep{lang2005does}. Finally, we plan to investigate whether the public's perceived lack of political power makes them more distrustful of institutions to govern AI. In a recent survey of the British public, a large majority felt that they were unable to influence AI development, and many felt tech companies and governments have disproportionate influence \citep{cave2019scary}. Research on public perceptions of AI is nascent but could have a broad impact as AI governance moves from the realm of abstract principles into the world of mass politics.

%
\bibliographystyle{abbrvnat}
\bibliography{references}

\begin{thebibliography}{39}
\providecommand{\natexlab}[1]{#1}
\providecommand{\url}[1]{\texttt{#1}}
\expandafter\ifx\csname urlstyle\endcsname\relax
  \providecommand{\doi}[1]{doi: #1}\else
  \providecommand{\doi}{doi: \begingroup \urlstyle{rm}\Url}\fi

\bibitem[{Ada Lovelace Institute}(2019)]{facevalue2019}
{Ada Lovelace Institute}.
\newblock Beyond face value: public attitudes to facial recognition technology.
\newblock Technical report, Ada Lovelace Institute, 2019.
\newblock URL \url{https://perma.cc/M4EG-N5MY}.

\bibitem[Aghion et~al.(2010)Aghion, Algan, Cahuc, and
  Shleifer]{aghion2010regulation}
P.~Aghion, Y.~Algan, P.~Cahuc, and A.~Shleifer.
\newblock Regulation and distrust.
\newblock \emph{The Quarterly Journal of Economics}, 125\penalty0 (3):\penalty0
  1015--1049, 2010.

\bibitem[Casey~Newton and Zelenko(2017)]{newton2017verge}
N.~S. Casey~Newton and M.~Zelenko.
\newblock The verge tech survey.
\newblock \emph{The Verge}, 2017.
\newblock URL \url{https://perma.cc/F75M-ZRYA}.

\bibitem[Caughey and Warshaw(2018)]{caughey2018policy}
D.~Caughey and C.~Warshaw.
\newblock Policy preferences and policy change: Dynamic responsiveness in the
  american states, 1936--2014.
\newblock \emph{American Political Science Review}, 112\penalty0 (2):\penalty0
  249--266, 2018.

\bibitem[Cave et~al.(2019)Cave, Coughlan, and Dihal]{cave2019scary}
S.~Cave, K.~Coughlan, and K.~Dihal.
\newblock Scary robots: examining public responses to ai.
\newblock In \emph{Proceedings of the 2019 AAAI/ACM Conference on AI, Ethics,
  and Society}, pages 331--337. ACM, 2019.

\bibitem[Cobb and Macoubrie(2004)]{macoubrie2004public}
M.~D. Cobb and J.~Macoubrie.
\newblock Public perceptions about nanotechnology: Risks, benefits and trust.
\newblock \emph{Journal of Nanoparticle Research}, 6\penalty0 (4):\penalty0
  395--405, 2004.

\bibitem[Doherty and Kiley(2019)]{doherty2019americans}
C.~Doherty and J.~Kiley.
\newblock Americans have become much less positive about tech companies’
  impact on the u.s.
\newblock Technical report, Pew Research Center, 2019.
\newblock URL \url{https://perma.cc/PHQ3-LD8D}.

\bibitem[Eurobarometer(2017)]{eurobarometer460}
Eurobarometer.
\newblock Special eurobarometer 460: Attitudes towards the impact of
  digitisation and automation on daily life.
\newblock Technical report, Eurobarometer, 2017.
\newblock URL \url{https://perma.cc/9FRT-ADST}.

\bibitem[Fischer and Wenger(2019)]{politicallyneutral2019}
S.-C. Fischer and A.~Wenger.
\newblock A politically neutral hub for basic ai research.
\newblock Technical report, Center for Security Studies (CSS) at ETH Zurich,
  2019.
\newblock URL \url{https://perma.cc/JA6M-RDUS}.

\bibitem[Floridi and Cowls(2019)]{floridi2019unified}
L.~Floridi and J.~Cowls.
\newblock A unified framework of five principles for ai in society.
\newblock \emph{Harvard Data Science Review}, 1\penalty0 (1), 6 2019.
\newblock \doi{10.1162/99608f92.8cd550d1}.
\newblock URL \url{https://hdsr.mitpress.mit.edu/pub/l0jsh9d1}.
\newblock https://hdsr.mitpress.mit.edu/pub/l0jsh9d1.

\bibitem[Funk(2017)]{funk2017}
C.~Funk.
\newblock Real numbers: Mixed messages about public trust in science.
\newblock \emph{Issues in Science and Technology}, 34\penalty0 (1), 2017.
\newblock URL \url{https://perma.cc/UF9P-WSRL}.

\bibitem[Funk and Kennedy(2016)]{funk2016new}
C.~Funk and B.~Kennedy.
\newblock The new food fights: U.s. public divides over food science.
\newblock Technical report, Pew Research Center, 2016.
\newblock URL \url{https://perma.cc/TJ2L-K9JU}.

\bibitem[Horowitz(2016)]{horowitz2016public}
M.~C. Horowitz.
\newblock Public opinion and the politics of the killer robots debate.
\newblock \emph{Research \& Politics}, 3\penalty0 (1), 2016.
\newblock URL \url{https://doi.org/10.1177/2053168015627183}.

\bibitem[Kahn and Ingram(2018)]{kahn2018americans}
C.~Kahn and D.~Ingram.
\newblock Americans less likely to trust facebook than rivals on personal data:
  Reuters/ipsos poll.
\newblock \emph{Reuters}, 2018.
\newblock URL \url{https://perma.cc/5KSL-2253}.

\bibitem[Lang and Hallman(2005)]{lang2005does}
J.~T. Lang and W.~K. Hallman.
\newblock Who does the public trust? the case of genetically modified food in
  the united states.
\newblock \emph{Risk Analysis: An International Journal}, 25\penalty0
  (5):\penalty0 1241--1252, 2005.

\bibitem[Lin and Green(2016)]{lin2016standard}
W.~Lin and D.~P. Green.
\newblock Standard operating procedures: A safety net for pre-analysis plans.
\newblock \emph{PS: Political Science \& Politics}, 49\penalty0 (3):\penalty0
  495--500, 2016.

\bibitem[Macoubrie(2006)]{macoubrie2006nanotechnology}
J.~Macoubrie.
\newblock Nanotechnology: public concerns, reasoning and trust in government.
\newblock \emph{Public Understanding of Science}, 15\penalty0 (2):\penalty0
  221--241, 2006.

\bibitem[Marques et~al.(2015)Marques, Critchley, and
  Walshe]{marques2015attitudes}
M.~D. Marques, C.~R. Critchley, and J.~Walshe.
\newblock Attitudes to genetically modified food over time: How trust in
  organizations and the media cycle predict support.
\newblock \emph{Public Understanding of Science}, 24\penalty0 (5):\penalty0
  601--618, 2015.

\bibitem[Mittelstadt(2019)]{mittelstadt2019ai}
B.~Mittelstadt.
\newblock Ai ethics--too principled to fail?
\newblock Forthcoming in \textit{Nature Machine Intelligence}, 2019.
\newblock URL \url{https://dx.doi.org/10.2139/ssrn.3391293}.

\bibitem[Molla(2018)]{molla2018facebook}
R.~Molla.
\newblock Facebook is the least-trusted major tech company.
\newblock \emph{Recode}, 2018.
\newblock URL \url{https://perma.cc/F75M-ZRYA}.

\bibitem[Nosek et~al.(2018)Nosek, Ebersole, DeHaven, and
  Mellor]{nosek2018preregistration}
B.~A. Nosek, C.~R. Ebersole, A.~C. DeHaven, and D.~T. Mellor.
\newblock The preregistration revolution.
\newblock \emph{Proceedings of the National Academy of Sciences}, 115\penalty0
  (11):\penalty0 2600--2606, 2018.

\bibitem[{OECD}(2019)]{oecdai}
{OECD}.
\newblock \emph{Artificial Intelligence in Society}.
\newblock 2019.
\newblock \doi{https://doi.org/https://doi.org/10.1787/eedfee77-en}.

\bibitem[Olofsson et~al.(2006)Olofsson, {\"O}hman, and
  Rashid]{olofsson2006attitudes}
A.~Olofsson, S.~{\"O}hman, and S.~Rashid.
\newblock Attitudes to gene technology: the significance of trust in
  institutions.
\newblock \emph{European Societies}, 8\penalty0 (4):\penalty0 601--624, 2006.

\bibitem[Pitlik and Kouba(2015)]{pitlik2015does}
H.~Pitlik and L.~Kouba.
\newblock Does social distrust always lead to a stronger support for government
  intervention?
\newblock \emph{Public Choice}, 163\penalty0 (3-4):\penalty0 355--377, 2015.

\bibitem[Satterfield et~al.(2009)Satterfield, Kandlikar, Beaudrie, Conti, and
  Harthorn]{satterfield2009anticipating}
T.~Satterfield, M.~Kandlikar, C.~E. Beaudrie, J.~Conti, and B.~H. Harthorn.
\newblock Anticipating the perceived risk of nanotechnologies.
\newblock \emph{Nature Nanotechnology}, 4\penalty0 (11):\penalty0 752, 2009.

\bibitem[Saxena et~al.(2019)Saxena, Huang, DeFilippis, Radanovic, Parkes, and
  Liu]{saxena2019fairness}
N.~A. Saxena, K.~Huang, E.~DeFilippis, G.~Radanovic, D.~C. Parkes, and Y.~Liu.
\newblock How do fairness definitions fare?: Examining public attitudes towards
  algorithmic definitions of fairness.
\newblock In \emph{Proceedings of the 2019 AAAI/ACM Conference on AI, Ethics,
  and Society}, pages 99--106. ACM, 2019.

\bibitem[Scott et~al.(2018)Scott, Inbar, Wirz, Brossard, and
  Rozin]{scott2018overview}
S.~E. Scott, Y.~Inbar, C.~D. Wirz, D.~Brossard, and P.~Rozin.
\newblock An overview of attitudes toward genetically engineered food.
\newblock \emph{Annual Review of Nutrition}, 38:\penalty0 459--479, 2018.

\bibitem[Siegrist(1999)]{siegrist1999causal}
M.~Siegrist.
\newblock A causal model explaining the perception and acceptance of gene
  technology 1.
\newblock \emph{Journal of Applied Social Psychology}, 29\penalty0
  (10):\penalty0 2093--2106, 1999.

\bibitem[Siegrist(2000)]{siegrist2000influence}
M.~Siegrist.
\newblock The influence of trust and perceptions of risks and benefits on the
  acceptance of gene technology.
\newblock \emph{Risk Analysis}, 20\penalty0 (2):\penalty0 195--204, 2000.

\bibitem[Siegrist et~al.(2005)Siegrist, Gutscher, and
  Earle]{siegrist2005perception}
M.~Siegrist, H.~Gutscher, and T.~C. Earle.
\newblock Perception of risk: the influence of general trust, and general
  confidence.
\newblock \emph{Journal of Risk Research}, 8\penalty0 (2):\penalty0 145--156,
  2005.

\bibitem[Siegrist et~al.(2007{\natexlab{a}})Siegrist, Cousin, Kastenholz, and
  Wiek]{siegrist2007public}
M.~Siegrist, M.-E. Cousin, H.~Kastenholz, and A.~Wiek.
\newblock Public acceptance of nanotechnology foods and food packaging: The
  influence of affect and trust.
\newblock \emph{Appetite}, 49\penalty0 (2):\penalty0 459--466,
  2007{\natexlab{a}}.

\bibitem[Siegrist et~al.(2007{\natexlab{b}})Siegrist, Keller, Kastenholz, Frey,
  and Wiek]{siegrist2007laypeople}
M.~Siegrist, C.~Keller, H.~Kastenholz, S.~Frey, and A.~Wiek.
\newblock Laypeople's and experts' perception of nanotechnology hazards.
\newblock \emph{Risk Analysis: An International Journal}, 27\penalty0
  (1):\penalty0 59--69, 2007{\natexlab{b}}.

\bibitem[Smith(2018)]{smith2018}
A.~Smith.
\newblock Public attitudes toward technology companies.
\newblock Survey report, Pew Research Center, 2018.
\newblock URL \url{https://perma.cc/KSN6-6FRW}.

\bibitem[Smith(2019)]{smith2019facial}
A.~Smith.
\newblock More than half of u.s. adults trust law enforcement to use facial
  recognition responsibly.
\newblock Technical report, Pew Research Center, 2019.
\newblock URL \url{https://perma.cc/FUV7-5BDJ}.

\bibitem[Smith and Anderson(2016)]{smith2017}
A.~Smith and M.~Anderson.
\newblock Automation in everyday life.
\newblock Technical report, Pew Research Center, 2016.
\newblock URL \url{https://perma.cc/WU6B-63PZ}.

\bibitem[West(2018)]{west2018divided}
D.~M. West.
\newblock Brookings survey finds divided views on artificial intelligence for
  warfare, but support rises if adversaries are developing it.
\newblock Survey report, Brookings Institution, 2018.
\newblock URL \url{https://perma.cc/3NJV-5GV4}.

\bibitem[West et~al.(2019)West, Whittaker, and Crawford]{west2019}
S.~M. West, M.~Whittaker, and K.~Crawford.
\newblock Discriminating systems: Gender, race and power in ai.
\newblock Technical report, AI Now Institute, 2019.
\newblock URL \url{https://perma.cc/85VP-HL8R}.

\bibitem[Whittlestone et~al.(2019)Whittlestone, Nyrup, Alexandrova, and
  Cave]{whittlestone2019role}
J.~Whittlestone, R.~Nyrup, A.~Alexandrova, and S.~Cave.
\newblock The role and limits of principles in ai ethics: towards a focus on
  tensions.
\newblock In \emph{Proceedings of the 2019 AAAI/ACM Conference on AI, Ethics,
  and Society}, pages 195--200. ACM, 2019.

\bibitem[Zhang(2019)]{zhang2019regulation}
B.~Zhang.
\newblock Public opinion lessons for {AI} regulation.
\newblock Technical report, Brookings Institution, 2019.
\newblock URL \url{https://perma.cc/B3FY-K6JH}.

\end{thebibliography}

\newpage

\appendix

\section{YouGov Sampling and Weights}

YouGov interviewed 2387 respondents who were then matched down to a sample of 2000 to produce the final dataset. The respondents were matched to a sampling frame on gender, age, race, and education. The frame was constructed by stratified sampling from the full 2016 American Community Survey (ACS) one-year sample with selection within strata by weighted sampling with replacements (using the person weights on the public use file).

The matched cases were weighted to the sampling frame using propensity scores. The matched cases and the frame were combined and a logistic regression was estimated for inclusion in the frame. The propensity score function included age, gender, race/ethnicity, years of education, and geographic region. The propensity scores were grouped into deciles of the estimated propensity score in the frame and post-stratified according to these deciles.

The weights were then post-stratified on 2016 U.S. presidential vote choice, and a four-way stratification of gender, age (four-categories), race (four-categories), and education (four-categories), to produce the final weight.

\section{Text of the Questions}

Below, we present the survey text as shown to respondents. The numerical codings are shown in parentheses following each answer choice. 

\subsection{Should AI and/or robots should be carefully managed}

\noindent Please tell me to what extent you agree or disagree with the following statement.

\noindent [Respondents were presented with one statement randomly selected from the list below.]

\begin{itemize}
\item AI and robots are technologies that require careful management.
\item AI is a technology that requires careful management.
\item Robots are technologies that require careful management.
\end{itemize}

\noindent ANSWER CHOICES:
\begin{itemize}
\item Totally agree (2)
\item Tend to agree (1)
\item Tend to disagree (-1)
\item Totally disagree (-2)
\item I don’t know
\end{itemize}

\subsection{Support for developing AI}

\noindent [All respondents were presented with the following prompt.]

\noindent Next, we would like to ask you questions about your attitudes toward artificial intelligence.

\noindent Artificial Intelligence (AI) refers to computer systems that perform tasks or make decisions that usually require human intelligence. AI can perform these tasks or make these decisions without explicit human instructions. Today, AI has been used in the following applications:

\noindent [Respondents were shown five items randomly selected from the list below.]

\begin{itemize}
\item Translate over 100 different languages
\item Predict one’s Google searches
\item Identify people from their photos
\item Diagnose diseases like skin cancer and common illnesses
\item Predict who are at risk of various diseases
\item Help run factories and warehouses
\item Block spam email
\item Play computer games
\item Help conduct legal case research
\item Categorize photos and videos
\item Detect plagiarism in essays
\item Spot abusive messages on social media
\item Predict what one is likely to buy online
\item Predict what movies or TV shows one is likely to watch online
\end{itemize}

\noindent QUESTION: How much do you support or oppose the development of AI?

\noindent ANSWER CHOICES:

\begin{itemize}
\item Strongly support (2)
\item Somewhat support (1)
\item Neither support nor oppose (0)
\item Somewhat oppose (-1)
\item Strongly oppose (-2)
\item I don’t know
\end{itemize}

\subsection{AI governance challenges}

\noindent We would like you to consider some potential policy issues related to AI. Please consider the following:

\noindent [Respondents were shown five randomly-selected items from the list below, one item at a time. For ease of comprehension, we include the shorten labels used in the figures in bold.]

\begin{itemize}
\item \textbf{Hiring bias}: Fairness and transparency in AI used in hiring: Increasingly, employers are using AI to make hiring decisions. AI has the potential to make less biased hiring decisions than humans. But algorithms trained on biased data can lead to lead to hiring practices that discriminate against certain groups. Also, AI used in this application may lack transparency, such that human users do not understand what the algorithm is doing, or why it reaches certain decisions in specific cases.
\item \textbf{Criminal justice bias}: Fairness and transparency in AI used in criminal justice: Increasingly, the criminal justice system is using AI to make sentencing and parole decisions. AI has the potential to make less biased hiring decisions than humans. But algorithms trained on biased data could lead to discrimination against certain groups. Also, AI used in this application may lack transparency such that human users do not understand what the algorithm is doing, or why it reaches certain decisions in specific cases.
\item \textbf{Disease diagnosis}: Accuracy and transparency in AI used for disease diagnosis: Increasingly, AI software has been used to diagnose diseases, such as heart disease and cancer. One challenge is to make sure the AI can correctly diagnose those who have the disease and not mistakenly diagnose those who do not have the disease. Another challenge is that AI used in this application may lack transparency such that human users do not understand what the algorithm is doing, or why it reaches certain decisions in specific cases.
\item \textbf{Data privacy}: Protect data privacy: Algorithms used in AI applications are often trained on vast amounts of personal data, including medical records, social media content, and financial transactions. Some worry that data used to train algorithms are not collected, used, and stored in ways that protect personal privacy.
\item \textbf{Autonomous vehicles}: Make sure autonomous vehicles are safe: Companies are developing self-driving cars and trucks that require little or no input from humans. Some worry about the safety of autonomous vehicles for those riding in them as well as for other vehicles, cyclists, and pedestrians.
\item \textbf{Ditigal manipulation}: Prevent AI from being used to spread fake and harmful content online: AI has been used by governments, private groups, and individuals to harm or manipulate internet users. For instance, automated bots have been used to generate and spread false and/or harmful news stories, audios, and videos.
\item \textbf{Cyber attacks}: Prevent AI cyber attacks against governments, companies, organizations, and individuals: Computer scientists have shown that AI can be used to launch effective cyber attacks. AI could be used to hack into servers to steal sensitive information, shut down critical infrastructures like power grids or hospital networks, or scale up targeted phishing attacks.
\item \textbf{Surveillance}: Prevent AI-assisted surveillance from violating privacy and civil liberties: AI can be used to process and analyze large amounts of text, photo, audio, and video data from social media, mobile communications, and CCTV cameras. Some worry that governments, companies, and employers could use AI to increase their surveillance capabilities.
\item \textbf{U.S.-China arms race}: Prevent escalation of a U.S.-China AI arms race: Leading analysts believe that an AI arms race is beginning, in which the U.S. and China are investing billions of dollars to develop powerful AI systems for surveillance, autonomous weapons, cyber operations, propaganda, and command and control systems. Some worry that a U.S.-China arms race could lead to extreme dangers. To stay ahead, the U.S. and China may race to deploy advanced military AI systems that they do not fully understand or can control. We could see catastrophic accidents, such as a rapid, automated escalation involving cyber and nuclear weapons.
\item \textbf{Value alignment}: Make sure AI systems are safe, trustworthy, and aligned with human values: As AI systems become more advanced, they will increasingly make decisions without human input. One potential fear is that AI systems, while performing jobs they are programmed to do, could unintentionally make decisions that go against the values of its human users, such as physically harming people.
\item \textbf{Autonomous weapons}: Ban the use of lethal autonomous weapons (LAWs): Lethal autonomous weapons (LAWs) are military robots that can attack targets without control by humans. LAWs could reduce the use of human combatants on the battlefield. But some worry that the adoption of LAWs could lead to mass violence. Because they are cheap and easy to produce in bulk, national militaries, terrorists, and other groups could readily deploy LAWs.
\item \textbf{Technological unemployment}: Guarantee a good standard of living for those who lose their jobs to automation: Some forecast that AI will increasingly be able to do jobs done by humans today. AI could potentially do the jobs of blue-collar workers, like truckers and factory workers, as well as the jobs of white-collar workers, like financial analysts or lawyers. Some worry that in the future, robots and computers can do most of the jobs that are done by humans today.
\item \textbf{Critical AI systems failure}: Prevent critical AI systems failures: As AI systems become more advanced, they could be used by the military or in critical infrastructure, like power grids, highways, or hospital networks. Some worry that the failure of AI systems or unintentional accidents in these applications could cause 10 percent or more of all humans to die.
\end{itemize}

\noindent QUESTION: In the next 10 years, how likely do you think it is that this AI governance challenge will impact large numbers of people in the U.S.?

\noindent ANSWER CHOICES:

\begin{itemize}
\item Very unlikely: less than 5\% chance (2.5\%)
\item Unlikely: 5-20\% chance (12.5\%)
\item Somewhat unlikely: 20-40\% chance (30\%)
\item Equally likely as unlikely: 40-60\% chance (50\%)
\item Somewhat likely: 60-80\% chance (70\%)
\item Likely: 80-95\% chance (87.5\%)
\item Very likely: more than 95\% chance (97.5\%)
\item I don’t know
\end{itemize}

\noindent QUESTION: In the next 10 years, how likely do you think it is that this AI governance challenge will impact large numbers of people around the world?

\noindent ANSWER CHOICES:

\begin{itemize}
\item Very unlikely: less than 5\% chance (2.5\%)
\item Unlikely: 5-20\% chance (12.5\%)
\item Somewhat unlikely: 20-40\% chance (30\%)
\item Equally likely as unlikely: 40-60\% chance (50\%)
\item Somewhat likely: 60-80\% chance (70\%)
\item Likely: 80-95\% chance (87.5\%)
\item Very likely: more than 95\% chance (97.5\%)
\item I don’t know
\end{itemize}

\noindent QUESTION: In the next 10 years, how important is it for tech companies and governments to carefully manage the following challenge?

\noindent ANSWER CHOICES:

\begin{itemize}
\item Very important (3)
\item Somewhat important (2)
\item Not too important (1)
\item Not at all important (0)
\item I don’t know
\end{itemize}

\subsection{Trust of actors to develop AI}

\noindent QUESTION: How much confidence, if any, do you have in each of the following to develop AI in the best interests of the public?

\noindent [Respondents were shown five items randomly selected from the list below. We included explainer text for actors not well known to the public; respondents could view the explainer text by hovering their mouse over the actor’s name. The items and the answer choices were shown in a matrix format.]

\begin{itemize}
\item The U.S. military
\item The U.S. civilian government
\item National Security Agency (NSA)
\item Federal Bureau of Investigation (FBI)
\item Central Intelligence Agency (CIA)
\item North Atlantic Treaty Organization (NATO);  explainer text for NATO: NATO is a military alliance that includes 28 countries including most of Europe, as well as the U.S. and Canada.
\item An international research organization (e.g., CERN); explainer text for CERN: The European Organization for Nuclear Research, known as CERN, is a European research organization that operates the largest particle physics laboratory in the world.
\item Tech companies
\item Google
\item Facebook
\item Apple
\item Microsoft
\item Amazon
\item A non-profit AI research organization (e.g., OpenAI); explainer text for OpenAI: Open AI is an AI non-profit organization with backing from tech investors that seeks to develop safe AI. University researchers
\end{itemize}

\noindent ANSWER CHOICES:

\begin{itemize}
\item A great deal of confidence (3)
\item A fair amount of confidence (2)
\item Not too much confidence (1)
\item No confidence (0)
\item I don’t know
\end{itemize}

\subsection{Trust of actors to manage AI}

\noindent QUESTION: How much confidence, if any, do you have in each of the following to manage the development and use of AI in the best interests of the public?

\noindent [Respondents were shown five items randomly selected from the list below. We included explainer text for actors not well known to the public; respondents could view the explainer text by hovering their mouse over the actor’s name. The items and the answer choices were shown in a matrix format.]

\begin{itemize}
\item U.S. federal government
\item U.S. state governments
\item International organizations (e.g., United Nations, European Union)
\item The United Nations (UN)
\item An intergovernmental research organization (e.g., CERN); explainer text for CERN: The European Organization for Nuclear Research, known as CERN, is a European research organization that operates the largest particle physics laboratory in the world.
\item Tech companies
\item Google
\item Facebook
\item Apple
\item Microsoft
\item Amazon
\item Non-government scientific organizations (e.g., AAAI); explainer text for AAAI: Association for the Advancement of Artificial Intelligence (AAAI) is a non-government scientific organization that promotes research in, and responsible use of AI.
Partnership on AI, an association of tech companies, academics, and civil society groups
\end{itemize}

\noindent ANSWER CHOICES:

\begin{itemize}
\item A great deal of confidence (3)
\item A fair amount of confidence (2)
\item Not too much confidence (1)
\item No confidence (0)
\item I don’t know
\end{itemize}

\FloatBarrier

\section{Additional Tables and Figures}

\begin{table}[h]
\centering
\begin{tabular}{p{5cm}p{2.5cm}}
\hline
Demographic subgroup & Unweighted sample size\\
\hline
Age 18-37 & 702\\

Age 38-53 & 506\\

Age 54-72 & 616\\

Age 73 and older & 176\\

Female & 1048\\

Male & 952\\

White & 1289\\

Non-white & 711\\

HS or less & 742\\

Some college & 645\\

College+ & 613\\

Not employed & 1036\\

Employed (full- or part-time) & 964\\

Income less than \$30K & 531\\

Income \$30-70K & 626\\

Income \$70-100K & 240\\

Income more than \$100K & 300\\

Prefer not to say income & 303\\

Republican & 470\\

Democrat & 699\\

Independent/Other & 831\\

Christian & 1061\\

No religious affiliation & 718\\

Other religion & 221\\

Not born-again Christian & 1443\\

Born-again Christian & 557\\

No CS or engineering degree & 1805\\

CS or engineering degree & 195\\

No CS or programming experience & 1265\\

CS or programming experience & 735\\
\hline
\end{tabular}

\caption{\label{tab:subgroups}Size of demographic subgroups}
\end{table}

\begin{table}[h]
\centering
\begin{tabular}{p{4.5cm}p{3cm}}
\hline
Variable & Coefficient (SE)\\
\hline

Age 38-53 & 0.11 (0.07)\\

Age 54-72 & 0.35 (0.06)***\\

Age 73 and older & 0.44 (0.07)***\\

Male & 0.02 (0.05)\\

Non-white & -0.01 (0.05)\\

Some college & 0.03 (0.07)\\

College+ & 0.15 (0.07)*\\

Employed (full- or part-time) & -0.09 (0.06)\\

Income \$30-70K & 0.09 (0.08)\\

Income \$70-100K & 0.13 (0.10)\\

Income more than \$100K & -0.01 (0.10)\\

Prefer not to say income & 0.04 (0.08)\\

Democrat & 0.13 (0.07)\\

Independent/Other & 0.14 (0.07)\\

No religious affiliation & -0.04 (0.06)\\

Other religion & -0.05 (0.08)\\

Born-again Christian & 0.07 (0.07)\\

CS or engineering degree & -0.35 (0.10)***\\

CS or programming experience & -0.01 (0.07)\\

Criminal justice bias & 0.05 (0.13)\\

Disease diagnosis & -0.06 (0.14)\\

Data privacy & 0.16 (0.13)\\

Autonomous vehicles & -0.07 (0.14)\\

Digital manipulation & -0.14 (0.15)\\

Cyber attacks & 0.05 (0.14)\\

Surveillance & <0.01 (0.15)\\

U.S.-China arms race & 0.04 (0.13)\\

Value alignment & -0.06 (0.13)\\

Autonomous weapons & 0.06 (0.14)\\

Technological unemployment & -0.12 (0.14)\\

Critical AI systems failure & -0.27 (0.15)\\

Intercept & 2.25 (0.11)***\\
\hline
$N$ = 10000 observations, 2000 respondents & $F$(259,1999) = 3.36; $p$-value: <0.001\\
\hline
\end{tabular}
\caption{\label{tab:aigovregsat}Results from a saturated regression predicting perceived issue importance using demographic variables, AI governance challenge, and interactions between the two types of variables; the coefficients for the interactions variables are not shown due to space constraints}
\end{table}

\begin{figure}[h]
\centering
\includegraphics[width=0.9\textwidth]{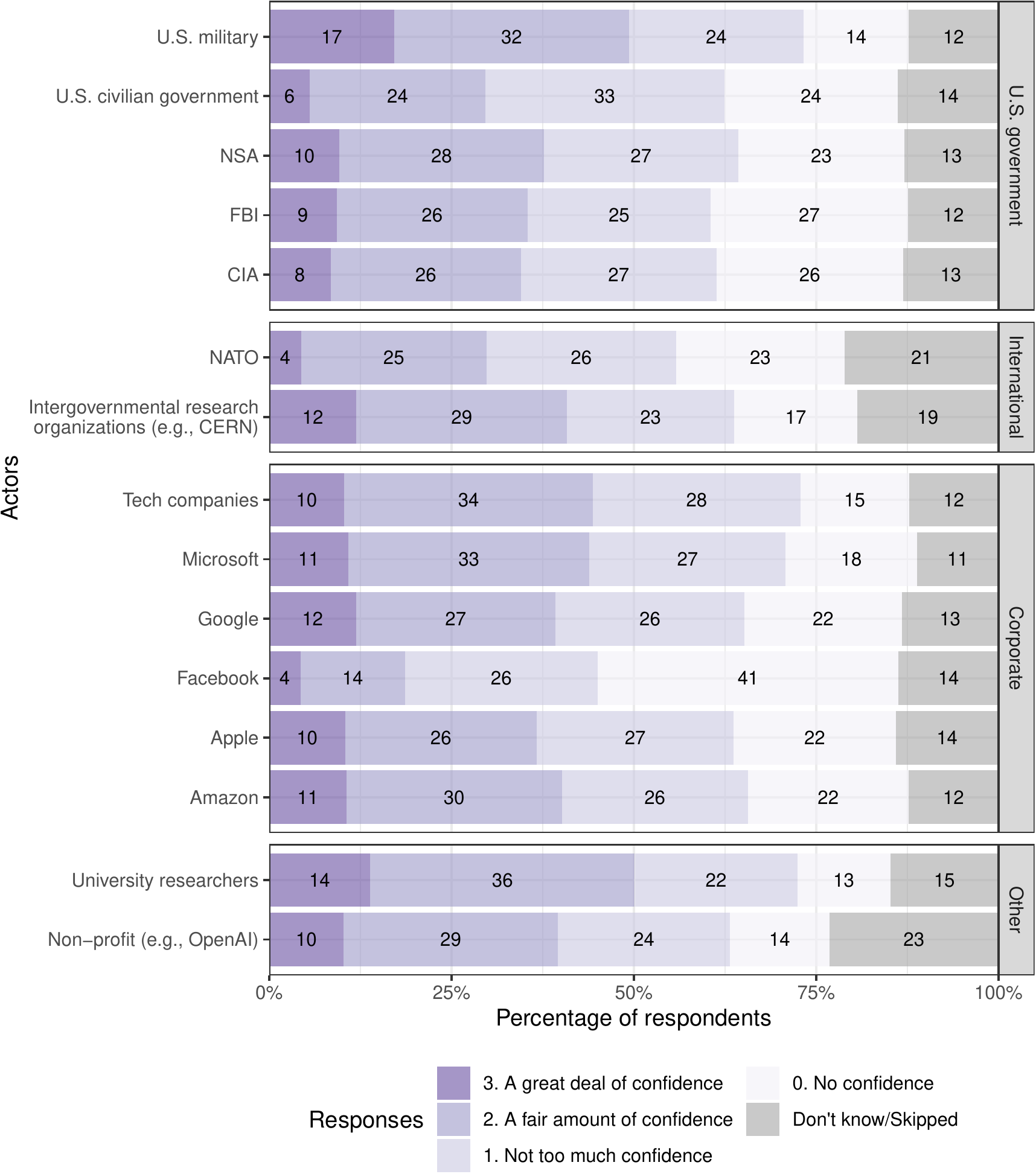}
\caption{Trust in various actors to develop AI: distribution of responses}
\label{fig:trustdev}
\end{figure}

\begin{figure}[h]
\centering
\includegraphics[width=0.9\textwidth]{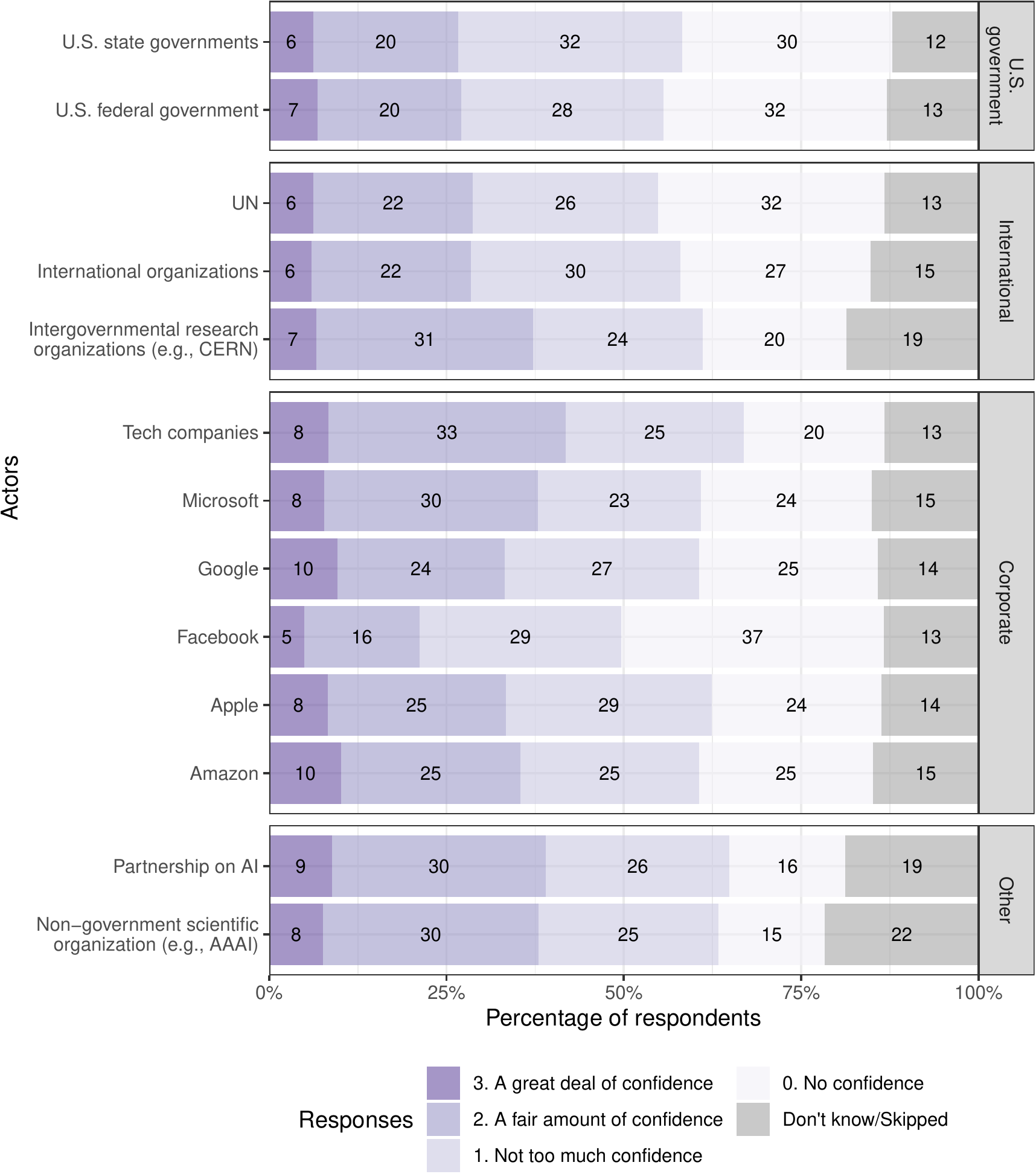}
\caption{Trust in various actors to manage AI: distribution of responses}
\label{fig:trustmanage}
\end{figure}

\begin{figure}[hbtp]
\centering
\includegraphics[width=0.95\textwidth]{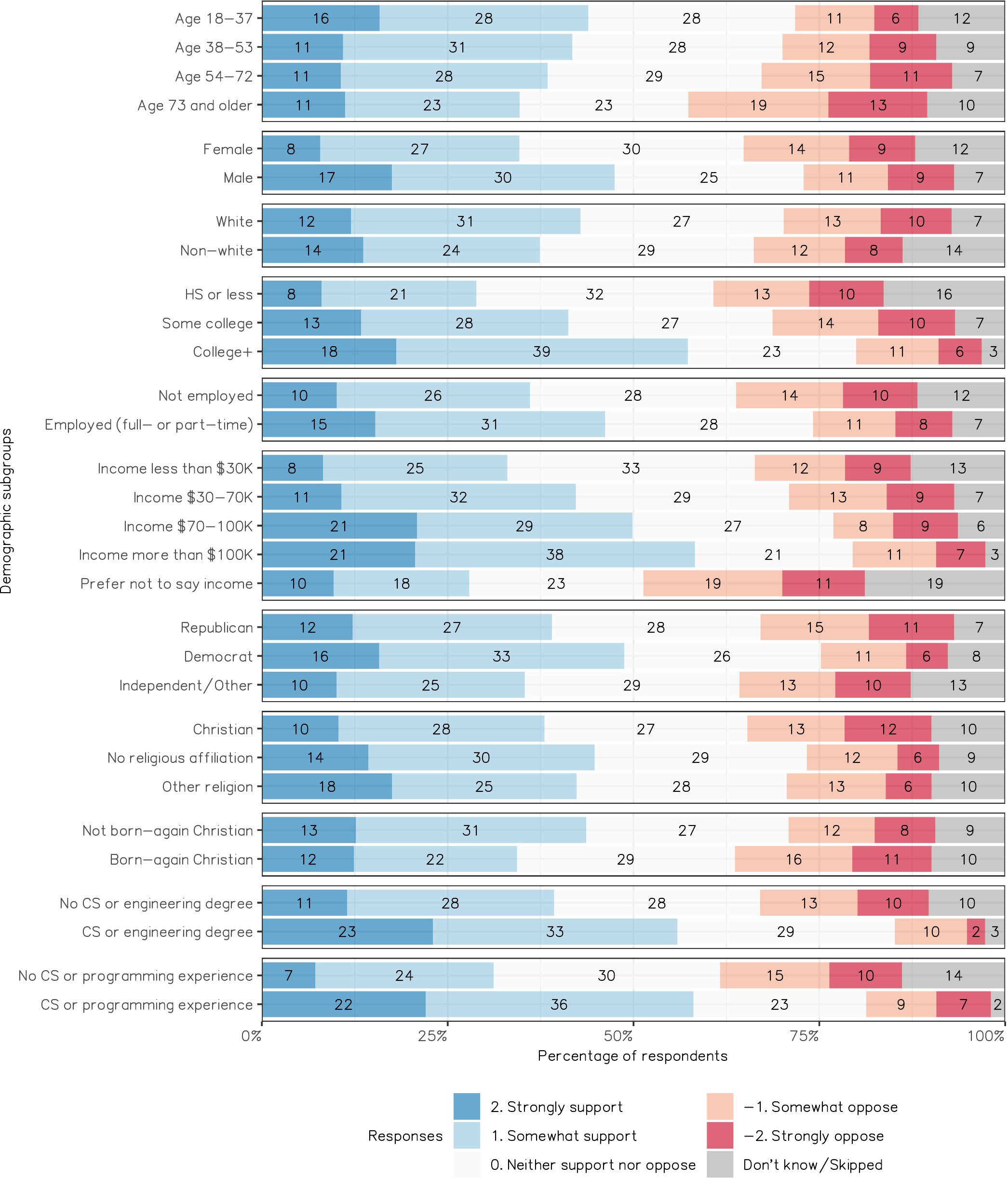}
\caption{Support for developing AI across demographic characteristics: distribution of responses}
\label{fig:supportdemo}
\end{figure}

\FloatBarrier

\end{document}